%% file: main.tex
\documentclass[aps,prb,twocolumn,superscriptaddress,longbibliography]{revtex4-2}
\include{header}
\usepackage{braket}
\usepackage[english]{babel}
\usepackage{amsmath, amssymb}
\usepackage{graphicx}
\usepackage{dcolumn}
\usepackage{bm}
\usepackage{float}
\usepackage{csvsimple}
\usepackage{nicefrac}
\usepackage{glossaries}
\usepackage{placeins}
\graphicspath{{Figures/}}
\usepackage{tikz}
\include{acronyms}
\usetikzlibrary{external}
\usepackage{color}

\include{tikzdefs}

\begin{document}
\title{Fusing matrix-product states and quantum Monte Carlo: reducing entanglement and sign problem at the same time}
\author{Gunnar Bollmark}
\affiliation{SUPA, Institute of Photonics and Quantum Sciences, Heriot-Watt University, Edinburgh EH14 4AS, United Kingdom}
\author{Sam Mardazad}
\affiliation{SUPA, Institute of Photonics and Quantum Sciences, Heriot-Watt University, Edinburgh EH14 4AS, United Kingdom}
\author{Johannes S. Hofmann}
\affiliation{Department of Condensed Matter Physics, Weizmann Institute of Science, Rehovot 76100, Israel}
\affiliation{Max-Planck-Institut f\"ur Physik komplexer Systeme, N\"othnitzer Strasse 38, 01187 Dresden, Germany}
\author{Adrian Kantian}
\affiliation{SUPA, Institute of Photonics and Quantum Sciences, Heriot-Watt University, Edinburgh EH14 4AS, United Kingdom}
\begin{abstract}
	Systems of correlated quantum matter can be a steep challenge to any would-be method of solution. 
	Matrix-product state (MPS)-based methods can describe 1D systems quasiexactly, but often struggle to retain sufficient bipartite entanglement to accurately approximate 2D systems already. 
	Conversely, Quantum Monte Carlo (QMC) approaches, based on sampling a probability distribution, can generally approximate 2D and 3D systems with an error that decays systematically with growing sampling size.
	However, QMC can suffer from the so-called sign problem, that makes the approach prohibitively costly for many systems of interest, such as repulsively interacting fermions away from commensurate densities and frustrated systems.
	In this article, we introduce a new hybrid approach, that combines auxiliary-field QMC (AFQMC) with MPS-based algorithms.
	This hybrid technique removes or reduces the sign problem (depending on the specific model) while also needing to retain much lower bipartite entanglement than brute-force application of a MPS-algorithm, without the use of uncontrolled approximations.
	We present two use-cases of the algorithm that would be challenging or impossible to address with any other approach, and quantify the extent of any remaining sign problem.
\end{abstract}
\maketitle

\section{\label{Sec::Introduction}Introduction}
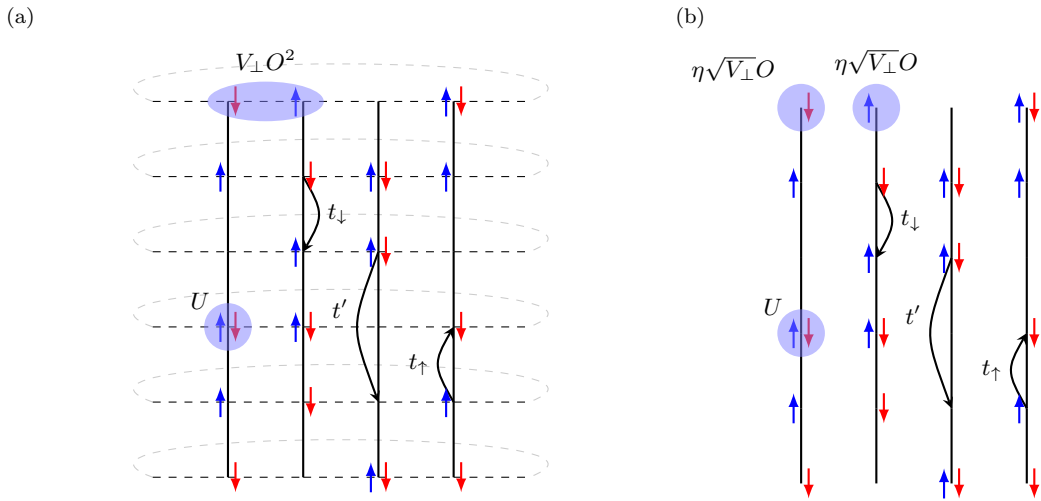
\begin{figure*}[!t]
	
	\subfloat[\label{Fig::SchemA}]{
		\ifthenelse{\boolean{buildtikzpics}}
		{
		\tikzsetnextfilename{SchematicModel}
		\tikzset{external/export next=true}
		\begin{tikzpicture}
			\foreach \y in {0,1,2,3,4,5}
			\draw[dashed,color=black!20] (-1,\y) to[out=160,in=20] (4,\y);
			\draw[->,thick] (1,4) to[out=-60,in=60,looseness=1.4, edge node={node[right,pos=0.5] {$t_\downarrow$}}](1,3);
			\draw[->,thick] (3,1) to[out=120,in=240,looseness=1.4, edge node={node[left,pos=0.5] {$t_\uparrow$}}](3,2);
			\draw[->,thick] (2,3) to[out=250,in=110,looseness=1.4, edge node={node[above left,pos=0.5] {$t^\prime$}}](2,1);
			\foreach \y in {0,1,2,...,5}
			\foreach \x in {-1,0,1,2,3}
			\draw[dashed] (\x,\y) -- (\x+1,\y);
			\foreach \x in {0,1,2,3}
			\foreach \y in {0,1,2,...,4}
			\draw[color=black!100,thick] (\x,\y) -- (\x,\y+1);
			\foreach \x in {0,1,2,3}{
				\edef\shift{1*\x}
				\foreach \yi[evaluate=\yi as \y using {Mod(\yi+\shift,6)}] in {1,2,4}{
					\draw[-latex, thick, color=blue] (\x-0.1, \y-0.2) -- (\x-0.1,\y+0.2) node (f\x_\y) {};}
				\edef\shift{1*\x+\x}
				\foreach \yi[evaluate=\yi as \y using {Mod(\yi+\shift,6)}] in {0,2,5}{
					\draw[-latex, thick, color=red] (\x+0.1, \y+0.2) -- (\x+0.1,\y-0.2) node (f\x_\y) {};
			}}
			\node[circle,fill=blue,color=blue!50, opacity=0.5,minimum size=18] (vertex) at (0,2) {};
			\node[above left=-0.1 and -0.1 of vertex] {$U$};
			\node[ellipse,minimum height=15, minimum width=44, color=blue!50, opacity=0.5, fill] (nnvertex) at (0.5,5) {};
			\node[above=0 of nnvertex] {$V_\perp O^2$};
		\end{tikzpicture}
	}{\includegraphics{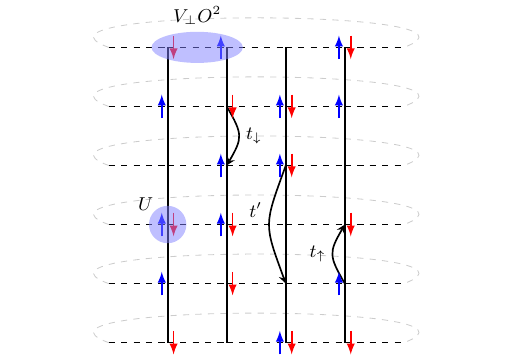}}
	}
	\subfloat[\label{Fig::SchemB}]{
		\ifthenelse{\boolean{buildtikzpics}}{
		\tikzsetnextfilename{SchematicDecouple}
		\tikzset{external/export next=true}
		\begin{tikzpicture}
			\draw[->,thick] (1,4) to[out=-60,in=60,looseness=1.4, edge node={node[right,pos=0.5] {$t_\downarrow$}}](1,3);
			\draw[->,thick] (3,1) to[out=120,in=240,looseness=1.4, edge node={node[left,pos=0.5] {$t_\uparrow$}}](3,2);
			\draw[->,thick] (2,3) to[out=250,in=110,looseness=1.4, edge node={node[above left,pos=0.5] {$t^\prime$}}](2,1);
			\foreach \x in {0,1,2,3}
			\foreach \y in {0,1,2,...,4}
			\draw[color=black!100,thick] (\x,\y) -- (\x,\y+1);
			\foreach \x in {0,1,2,3}{
				\edef\shift{1*\x}
				\foreach \yi[evaluate=\yi as \y using {Mod(\yi+\shift,6)}] in {1,2,4}{
					\draw[-latex, thick, color=blue] (\x-0.1, \y-0.2) -- (\x-0.1,\y+0.2) node (f\x_\y) {};}
				\edef\shift{1*\x+\x}
				\foreach \yi[evaluate=\yi as \y using {Mod(\yi+\shift,6)}] in {0,2,5}{
					\draw[-latex, thick, color=red] (\x+0.1, \y+0.2) -- (\x+0.1,\y-0.2) node (f\x_\y) {};
			}}
			\node[circle,fill=blue,color=blue!50, opacity=0.5,minimum size=18] (vertex) at (0,2) {};
			\node[above left=-0.1 and -0.1 of vertex] {$U$};
			\node[ellipse,minimum size=18, color=blue!50, opacity=0.5, fill] (nnvertex) at (0,5) {};
			\node[above left=0 of nnvertex] {$\eta\sqrt{V_\perp}O$};
			\node[circle,minimum size=18, color=blue!50, opacity=0.5, fill] (rnnvertex) at (1,5) {};
			\node[above=0 of rnnvertex] {$\eta\sqrt{V_\perp}O$};
		\end{tikzpicture}
	}{\includegraphics{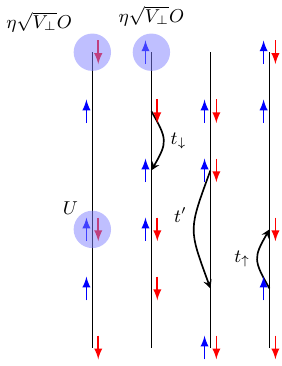}}
	}
	\caption{Figure~\ref{Fig::SchemA} shows the model considered in \cref{Eq::Hamiltonian}. Notably the transverse direction~(- - -) is always periodic whereas the longitudinal one~(-----) depends on the algorithm in use. Figure~\ref{Fig::SchemB} shows the system decoupled into \gls{1D} sub-units using \gls{HS} fields ($\eta$).}
\end{figure*}
Quantitative and unbiased numerics for correlated quantum systems constitute one of the most challenging areas of theoretical physics.
Such methods are indispensable when the aim is to predict which of several competing many-body states a complex material or model realizes at low temperatures~\cite{Georges2004,Bouillot2011,Dupont2020b}, or to aid in resolving fundamental problems such as the origin of unconventional (high-$T_c$) superconductivity in a variety of minimal models derived from real materials~\cite{Qin2020}
Yet, for these and many other problems of interest, the existing classes of algorithms often reach the limits of their usefulness well below the system sizes and/or above temperatures that would be required to observe the studied system's thermodynamic properties.
Techniques based on the \gls{MPS}, which are not explicitly biased towards any one order, can approximate a large variety of \gls{1D} systems with very high accuracy~\cite{Schollwock2005,Schollwock2011,Paeckel2019}.
However, this ability of \gls{MPS} algorithms is predicated on the bipartite entanglement of such systems being near-constant in system size, as the \gls{MPS}-approach achieves its high accuracy by retaining only the leading contributions to entanglement.
Once there are too many of such leading components, i.e. large Schmidt-coefficients~\cite{Schollwock2011}, as e.g. in many time-evolving \gls{1D} non-equilibrium states, or in \gls{2D} and \gls{3D} systems already at equilibrium, the \gls{MPS}-approximation can rapidly stop being quasi-exact or even useful.

Another class of unbiased algorithms, the \gls{QMC}-based methods, are not explicitly limited by the bipartite entanglement of a system.
Whether it is the \gls{AFQMC} approach~\cite{Blankenbecler1981,Assaad2022}, the continuous-time methods~\cite{Gull2011}, or the stochastic series expansion~\cite{Sandvik2019}, in every case the initial quantum many-body problem is mapped to a classical probability distribution function.
Ergodic sampling of this distribution leads to the stochastic error on any observables systematically shrinking with the number of samples.
For a great range of \gls{2D} and \gls{3D} systems, the \gls{QMC}-approach is the current de-facto gold standard for unbiased numerical first-principles calculations~\cite{LeBlanc2015,Schattner2016,Tang2018,Hofmann2022}, just as \gls{MPS}-algorithms are for \gls{1D} quantum systems.
Yet, when the quantum-to-classical mapping yields distribution functions that are not positive-definite, the \gls{QMC} approaches struggle against the so-called sign problem - the weight with which a specific configuration appears can now either be positive or negative.
This generally results in the statistical error of observables growing exponentially in system size and inverse temperature~(some exceptions lead to power-law growth~\cite{Zhang2022}). This, in turn, must be counteracted by an exponentially increasing, and thus rapidly infeasible, number of independent samples.
In this way, the sign problem strongly inhibits the study of many systems and phenomena of acute interest and importance. 
This challenge appears in many QMC-based works on high-$T_c$ superconductivity and it's competition with other correlated magnetic and insulating phases, to a wide variety of (potential) quantum spin liquids, as well as many other topics.

%
For \gls{MPS}-based approaches, bipartite subsystem entanglement will be at least proportional to subsystem surface area in practically any \gls{2D}/\gls{3D} system of interest~\cite{Stoudenmire2012}. This leads to a similar exponential resource requirement as for \gls{QMC}-based approaches.
These impasses have given rise to a number of hybrid and approximative variants of both the \gls{MPS} and \gls{QMC} approach.
On the \gls{MPS}-side, PEPS~\cite{Cirac2021} and other classes of tensor-network variational ansatz-states~\cite{Jiang2008} represent an extension of the original \gls{MPS}-concept to \gls{2D} and even \gls{3D} systems, sometimes as the basis of variational \gls{QMC}-algorithms~\cite{Sandvik2007,Schuch2008,Wang2011}.
Another approach, which is tailored to near-\gls{1D} systems in \gls{2D} and \gls{3D}, i.e. many weakly coupled \gls{1D} systems arrayed in parallel, hybridizes the \gls{MPS}-approach with \gls{MF} techniques.
First employed on spin systems~\cite{Klanjsek2008,Bouillot2011}, then bosons~\cite{Bollmark2020} and recently extended to fermions~\cite{Bollmark2023,Marten2023}, this technique too can treat very large systems.
However, the utility of these methods in strongly controlling or circumventing the growth of bipartite subsystem entanglement comes with the drawback that they will often be uncontrolled approximations to the original problem. 
Furthermore, they can also implicitly or explicitly end up biased towards some particular order(s), though the \gls{MPS}+\gls{MF} approach can address multiple competing ordered states on an equal footing~\cite{Bollmark2023a}, thus at least reducing the risk of bias.
On the \gls{QMC}-side, one can attempt to abate the sign-problem without giving up the exactness of the AFQMC-approach, as in variational AFQMC and the use of Lefschetz thimbles~\cite{Ulbyshev2019,Sorella2022} 
But, so far, the complete elimination of the sign problem can be achieved by deliberately removing negative-weight configurations, such as in constrained-path auxiliary-field \gls{QMC} (CP-AFQMC)~\cite{He2019}, but potentially again at the cost of uncontrolled approximations with the attendant risk of implicit bias.
A natural question in this context is thus whether a hybrid-algorithm exists which partly or wholly breaks free from the limitations of the pure \gls{MPS}- and \gls{QMC}-approaches, and which is simultaneously free from uncontrolled approximations.

In this article we present just such an algorithm, which uses a highly efficient \gls{MPS} representation for the different \gls{1D} slices making up a \gls{2D} system, while an \gls{AFQMC}-algorithm efficiently treats the coupling in-between these \gls{1D} slices.
We exhibit systems where our hybrid algorithm seemingly completely eliminates sign problems that quickly render a standard \gls{AFQMC}-treatment inefficient. Additionally, the new algorithm requires far more modest bipartite entanglement than would be present in a brute-force \gls{MPS}-treatment of the full \gls{2D} system.
Furthermore, we quantify the remaining sign problem potentially present in our hybrid algorithm, by studying the physics of mass-imbalanced lattice-fermions with repulsive interactions, doped away from half-filling. 
Such a system would not be possible to solve with \gls{AFQMC} at the system sizes and temperatures that our new algorithm can address.

We note that very recently \gls{MPS}-based techniques have started to be used to solve quantum impurity problems based on continuous-time perturbative expansions~\cite{NunezFernandez2022,Erpenbeck2023}, which are usually treated with \gls{QMC}-approaches.
Rather than hybridizing \gls{QMC}- and \gls{MPS}-approaches, this alternative paradigm replaces the Monte-Carlo sampling outright by approximating Feynman diagrams via a \gls{MPS}-representation.
However, in its current practical implementation, it would also be grouped with the effectively heuristic approaches listed above.  

The paper is structured as follows: In \cref{Sec::Model} we define and motivate the models with which the new method is developed. \cref{Sec::Method} derives the algorithm used to obtain a solution, the results of which are presented in \cref{Sec::Results}. Finally, the new method's properties are discussed in \cref{Sec::Discussion} and potential future applications are explored in \cref{Sec::Outlook}.

\section{\label{Sec::Model}Model}
The aim of the present work is to hybridize \gls{MPS}- and \gls{AFQMC}-techniques into an algorithm capable of treating \gls{2D} correlated lattice models without uncontrolled approximations, which combines the strengths of both approaches, while mitigating their respective limitations.
In particular, we aim to reduce or eliminate the sign problem that the pure \gls{AFQMC}-algorithms have for many models of \gls{2D} lattice fermions, while also reducing the bipartite entanglement that needs to be tracked by the \gls{MPS}-component of the algorithm, as compared to a brute-force \gls{MPS}-only calculation.
Thus, we focus on models which are difficult to impossible to solve with either \gls{AFQMC}- or \gls{MPS}-techniques alone, or with any other approach. 
For concreteness, we consider a class of models schematically represented in \cref{Fig::SchemA} which describes a structure of interacting \gls{1D} systems with no single-particle tunneling terms acting in-between them.
It is defined by $L_y$ copies of chains with next-to-nearest neighbour hopping arrayed in parallel - each ladder having $L_x$ sites overall - on which interacting spin-$1/2$ fermions move:
\begin{widetext}
	\begin{multline}\label{Eq::Hamiltonian}
		\hat{H} = \sum_{j=1}^{L_y}\Bigg[-\sum_{i\sigma}t^{\nodagger}_\sigma c^\dagger_{i,j,\sigma} c^\nodagger_{i+1,j,\sigma}+\text{h.c.} -t'\sum_{i\sigma} c^\dagger_{i,j,\sigma} c^\nodagger_{i+2,j,\sigma}+\text{h.c.} + \\U\sum_{i} n^{\nodagger}_{i,j,\uparrow}n^{\nodagger}_{i,j,\downarrow}
		+ \mu\sum_in^{\nodagger}_{i,j}+\sum_k\sum_i V^{(k)}_\perp \hat{W}^{(k)\nodagger}_{i,j}\hat{W}^{(k)\dagger}_{i,j+1}+\text{h.c.}\Bigg] =\sum_{j=1}^{L_y}\hat{H}^0_{1\text{D},j}+\sum_{j=1}^{L_y}\hat{H}_{\perp,j},
	\end{multline}
\end{widetext}
where
\begin{align}
	\lbrace &c_{i,j,\sigma}^\nodagger,c_{i^\prime,j^\prime,\sigma^\prime}^\dagger\rbrace=\delta_{i,i^\prime}\delta_{j,j^\prime}\delta_{\sigma,\sigma^\prime}, \\
	&n^\nodagger_{i,j,\sigma}=c^\dagger_{i,j,\sigma}c^\nodagger_{i,j,\sigma}, \\
	&n^\nodagger_{i,j} = \sum_\sigma n^\nodagger_{i,j,\sigma},
\end{align}
form common definitions of a fermionic system.
Here, $c_{i,j,\sigma}$ annihilates a fermion of spin $\sigma=\uparrow,\downarrow$ on site $i$ of the $j$th 1D system, and analogously for all other operators.
Additionally, we introduced the definition
\begin{equation}
	\hat{H}_{\perp,j} = \sum_k\sum_iV^{(k)}_\perp \hat{W}^{(k)\nodagger}_{i,j}\hat{W}^{(k)\dagger}_{i,j+1},
\end{equation}
for the transverse coupling Hamiltonian, and the operators $\hat{W}^{(k)\dagger}_{i,j}$, $\hat{W}^{(k)\nodagger}_{i,j}$ are different for different models, and are specified below.
Here, we assume that transverse coupling connects neighbouring sites, located on the $j$th and $j+1$th \gls{1D} systems respectively, but our algorithm is straightforwardly generalized to more long-ranged coupling between \gls{1D} systems.
All the remaining terms of \cref{Eq::Hamiltonian}, which represent the internal terms of the various chains, are stored in $\hat{H}^0_{1\text{D},j}$. 
The operators $\hat{W}^{(k)\nodagger}$ are quadratic in the fermionic creation and annihilation operators such that $\hat{H}_\perp,j$ is quartic. 

Each \gls{1D} system is represented by an extended \gls{1D} Hubbard model with a next-to-nearest neighbor hopping $t'$~(making each of these systems equivalent to a two-leg triangular ladder). 
Typically, the longer-range terms (i.e. $V$ and $t'$) will be weaker than the shorter-range ones.

In this work, we will consider two different sets for $\hat{W}^{(k)\nodagger}_{i,j}$: a repulsive interaction and an effective pair-hopping interaction in-between neighbouring \gls{1D} systems. 
The first interaction model to be studied is given by a single interaction term per site and per \gls{1D} system:
\begin{equation}\label{Eq::Repulsive}
	\hat{W}^{(1)\nodagger}_{i,j} = \hat{W}^{\nodagger}_{i,j} = n^{\nodagger}_{i,j},
\end{equation}
which simply corresponds to adjacent particles on neighbouring \gls{1D} systems repulsing each other. 
For this model, the on-site interaction will be kept strong and repulsive at $U=8t$ to realize the physics of strongly correlated fermions. 
Additionally, the longer-range hopping is introduced to exacerbate the sign problem and keep the fermionic nature of the model: With no such term the Hamiltonian is amenable to bosonic algorithms.

The latter interaction considered in this paper has been studied extensively using the \gls{MPS}+\gls{MF} algorithm for fermions co-developed by the authors~\cite{Bollmark2023,Marten2023} and is given by
\begin{align}\label{Eq::EffHopPairCreat}
	\hat{W}^{(1)\nodagger}_{i,j} &= c^{\dagger}_{i,j,\uparrow}c^{\dagger}_{i,j,\downarrow}, \\
	\label{Eq::EffHopPairAnn}
	\hat{W}^{(2)\nodagger}_{i,j} &= c_{i,j,\downarrow}c_{i,j,\uparrow}, \\
	\label{Eq::EffHopUp}
	\hat{W}^{(3)\nodagger}_{i,j} &= n^{\nodagger}_{i,j,\uparrow}, \\
	\label{Eq::EffHopDown}
	\hat{W}^{(4)\nodagger}_{i,j} &= n^{\nodagger}_{i,j,\downarrow}.
\end{align}
This interaction represents tunneling between chains where the spin gap dominates inter-chain tunneling~\cite{Bollmark2023}. 
In order to generate such a spin gap we study a model with strong attractive interactions $U=-10t$. 
For some parameter-regimes, such as half-filling, or mass-balanced doped systems, there are sign-problem-free pure-\gls{QMC} approaches~\cite{Huffman2014,Liu2015}. 
Conversely, for mass-imbalanced systems, $t_\uparrow\not=t_\downarrow$, and doping the system away from half filling there is a significant sign-problem.
So far, study of such systems has been limited to qualitative analytics~\cite{Cazalilla2005}.

Simulating low temperatures and large system sizes with standard \gls{QMC}-approaches for both the above models will inevitably create a sign problem that cannot be managed with available resources. 
Likewise, a brute-force \gls{MPS}-approach will have to deal with the need for resources growing exponentially with system size.
In the next section we present a method to partially circumvent this issue.
\section{\label{Sec::Method}Method}
In the following section we show how to combine \gls{AFQMC}- and \gls{MPS}-techniques for calculating the finite-temperature states of the Hamiltonian in Eq.~\eqref{Eq::Hamiltonian}. 
The way we combine both approaches is designed to avoid common problems that in existing \gls{AFQMC}- or \gls{MPS}-codes would encounter on their own.

\subsection{Auxiliary Field derivation}
While the Hamiltonian in \cref{Eq::Hamiltonian} has a \gls{2D} structure it is suitable to consider it a \gls{Q1D} model due to being composed of \gls{1D} systems connected only by an interaction and not by hopping.
While the entire model can be treated with \gls{AFQMC} this method suffers from the sign problem when no symmetry protects against it such as for the case of repulsive interactions at half-filling on a bipartite lattice~\cite{Li2019}. 
Moreover, \gls{MPS}-based methods are no complete solution as they are limited in how large \gls{2D} systems they can deal with~\cite{Schollwock2011}. 
In order to address both of these issues we consider a hybrid between the two approaches.

This hybrid strategy starts from the partition function
\begin{equation}\label{basepartit}
	\mathcal{Z} = \text{Tr}\left[e^{-\beta \hat{H}}\right] = \text{Tr}\left[\prod^{\beta/\mathrm{d}\tau}e^{-\mathrm{d}\tau \left[ (\sum_{i=1}^{L_y}\hat{H}_{1\text{D},i})+\hat{V}_\perp\right]}\right].
\end{equation}
Here, we have introduced 
\begin{align}\label{Eq::Newhamsplit}
       \nonumber \hat{H}_{1\text{D},i} &:= \hat{H}^0_{1\text{D},i} + \{\mbox{counter-terms}\},\\
	\hat{V}_\perp &:= \sum_{j=1}^{L_y} \hat{H}_{\perp,j} + \{\mbox{counter-terms}\}.
\end{align}
These counter-terms are chosen with two objectives in mind.
The first is that overall $\hat{H}$ stay invariant (up to irrelevant constants), i.e.
\begin{equation}\label{Eq::Splitidentity}
 \sum_{i=1}^{L_y} \hat{H}_{1\text{D},i} + \hat{V}_\perp =  \sum_{i=1}^{L_y} \hat{H}^0_{1\text{D},i} + \sum_{j=1}^{L_y} \hat{H}_{\perp,j} = \hat{H}.
\end{equation}
The second objective in choosing counter-terms is for the global coupling operator $\hat{V}_\perp$ attaining the property
\begin{equation}\label{Eq::Splitform}
	\hat{V}_\perp = \sum_n^{N_V}\hat{V}_{\perp,n} = \sum_n^{N_V}\hat{O}_n^2.
\end{equation}
Here, $n$ indexes all the individual terms $\hat{V}_{\perp,n}$ contained in $\hat{V}_\perp$, and the total number of all terms is denoted $N_V$.
Just as in the original inter-system coupling Hamiltonian $\sum_j  \hat{H}_{\perp,j}$, each term in \cref{Eq::Splitform} connects two sites on two different chains.
Depending on the specifics of the concrete model to be simulated, any such link may appear multiple times in the sum over $n$, each time with different operators acting on it, as is the case in e.g. the model treated in~\cref{SubSec::Results_SignProb}.
Thus, in general $n$ will be a composite index, comprised of the link $(i,j)-(i',j')$ (with $j\neq j'$), and a counter running over all the different products of operators acting on that link.

We exploit - and our algorithm demands from any model to be simulated with it - that all the operators $\hat{O}_n$ introduced in \cref{Eq::Splitform} be quadratic Hermitian operators. 
For any operator $\hat{V}_{\perp,n}$ respecting \cref{Eq::Splitform}, an identity utilizing Gauss-Hermite quadrature may be employed, also called \gls{HS} decoupling,~\cite{Assaad2022,AssaadCB2022}
\begin{equation}\label{gausshermite}
	e^{\mathrm{d}\tau \hat{V}_{\perp,n}} = \frac{\sqrt{\pi}}{4}\sum_{l=\pm1,\pm2}\gamma_le^{\sqrt{\mathrm{d}\tau}\eta_l\hat{O}_n}+\mathcal{O}(\mathrm{d}\tau^4),
\end{equation}
where $\eta_l$ are the auxiliary fields and $\gamma_l$ the associated integration prefactors given by
\begin{align}
	&\eta_l = \begin{cases}
		\pm\sqrt{2(3-\sqrt{6})} & \text{if } l=\pm1, \\
		\pm\sqrt{2(3+\sqrt{6})} & \text{if } l=\pm2,
	\end{cases} \\
	&\gamma_l = \begin{cases}
		1+\frac{\sqrt{6}}{3} & \text{if } l=\pm1, \\
		1-\frac{\sqrt{6}}{3} & \text{if } l=\pm2.
	\end{cases}
\end{align}
Furthermore, we require that the $\hat{O}_n$ quadratic operators are themselves composed of operators which belong only to a single \gls{1D} system:
\begin{equation}
	\hat{O}_n = \sum^{L_y}_ia^{(n)}_i\hat{o}_i,
\end{equation}
with arbitrary prefactors $a^{(n)}\in \mathbb{C}$, and where each $\hat{o}_i$ acts on the $i$-th \gls{1D}-system only. 
This allows the simplification of \cref{basepartit}
\begin{widetext}
\begin{multline}\label{splitpartit}
	\mathcal{Z} = \sum_{\left\{l_{\tau,n}\right\}}\text{Tr}\left[\prod^{\beta/\mathrm{d}\tau}_\tau e^{-\mathrm{d}\tau\sum_i\hat{H}_{1\text{D},i}}\prod_{n}^{N_V}\gamma_{\tau,n}e^{\sqrt{\mathrm{d}\tau}\eta_{\tau,n}\hat{O}_n}\right]+\mathcal{O}(\mathrm{d}\tau^2) \\ 
	= \sum_{\left\{l_{\tau,n}\right\}}\text{Tr}\left[\prod^{\beta/\mathrm{d}\tau}_\tau e^{-\mathrm{d}\tau\sum_i\hat{H}_{1\text{D},i}}\prod_{n}^{N_V}\gamma_{\tau,n}\prod_i^{L_y}e^{\sqrt{\mathrm{d}\tau}\eta_{\tau,n}a^{(n)}_i\hat{o}_i}\right]+\mathcal{O}(\mathrm{d}\tau^2) \\ 
	= \sum_{\mathcal{C}}\left(\prod_\tau^{\beta/d\tau}\prod_n^{N_V}\gamma_{\tau,n}\right)\text{Tr}\left[\prod_i^{L_y}\prod^{\beta/\mathrm{d}\tau}_\tau e^{-\mathrm{d}\tau\hat{H}_{1\text{D},i}}\prod_{n}^{N_V}e^{\sqrt{\mathrm{d}\tau}\eta_{\tau,n}a^{(n)}_i\hat{o}_i}\right]+\mathcal{O}(\mathrm{d}\tau^2),
\end{multline}
\end{widetext}
where we have used the convention $\eta_{\tau,n} = \eta_{l_{\tau,n}}$, similarly for $\gamma_{\tau,n}$ and the shorthand $\mathcal{C}=\left\{l_{\tau,n}\right\}$. 
In the last equality we have utilized the fact that operators belonging to different \gls{1D} systems will commute provided that they are composed of an even number of fermion operators. 
As a consequence, the trace is to be performed over an operator which lives in a product space of \gls{1D} systems:
\begin{equation}\label{Eq::partit_final}
	\mathcal{Z}=\sum_{\mathcal{C}}\Gamma_\mathcal{C}\text{Tr}\left[\bigotimes_i^{L_y} \hat{X}_{1\text{D},i}\right] = \sum_{\mathcal{C}}\Gamma_\mathcal{C}\prod_i^{L_y}\text{Tr}_{1\text{D}}\left[ \hat{X}_{1\text{D},i}\right],
\end{equation}
where
\begin{align}
	&\Gamma_\mathcal{C} = \left(\prod_\tau^{\beta/\mathrm{d}\tau}\prod_n^{N_V}\gamma_{\tau,n}\right),\\
	&\hat{X}_{1\text{D},i} = \prod^{\beta/\mathrm{d}\tau}_\tau e^{-\mathrm{d}\tau\hat{H}_{1\text{D},i}}\prod_{n}^{N_V}e^{\sqrt{\mathrm{d}\tau}\eta_{\tau,n}a^{(n)}_i\hat{o}_i}.
\end{align}
Ultimately, the partition function separates into a product of \gls{1D} traces with the cost of summing over a large space of configurations which is schematically visualized in \cref{Fig::SchemB}.
At this point, every non-quadratic operator in $\hat{H}_{1\text{D}}$ could be split using Gauss-Hermite quadrature which would yield the standard \gls{AFQMC} algorithm~\cite{Assaad2022,AssaadCB2022,Hao2016,Shiwei2019}. 
Instead, we will compute the product of \gls{1D} traces making up the partition function using numerically quasi-exact methods.

\subsection{Use of MPS time evolution}
Given the structure of \cref{Eq::partit_final} for the models we study here, the \gls{1D} partition functions, which are now functions of the \gls{HS}-variables to be stochastically sampled, are amenable to \gls{MPS}-based numerical methods, employing purification of states~\cite{Schollwock2011}:
\begin{equation}
	\text{Tr}\left[\hat{X}_{1\text{D}}\right] = \bra{\psi_0}\prod^{\beta/\mathrm{d}\tau}_\tau e^{-\mathrm{d}\tau\hat{H}_{1\text{D},i}}\prod_{n}^{N_V}e^{\sqrt{\mathrm{d}\tau}\eta_{\tau,n}a^{(n)}_i\hat{o}_i}\ket{\psi_0},
\end{equation}

where $\psi_0$ denotes the purification of the thermal density matrix at infinite temperature. 
The cost of this treatment is that each operator now acts both in the usual original Hilbert space but also as an identity in an auxiliary product space.

In this manner, computing traces for each term in the configuration series of \cref{Eq::partit_final} is equivalent to performing imaginary time evolution on \gls{1D} models. 
\gls{MPS}-based numerical methods are well placed to perform such calculations. 
There are multiple methods used to perform such time evolution~\cite{Paeckel2019}. 
In principle, any one time evolution method or mix thereof would be possible to use for this problem, and thus we can choose a combination yielding the best performance.

One benefit of this approach is that we avoid having to \gls{HS}-decouple the strong on-site interaction term. 

This could have implications on the severity of the sign problem. 
From the outset we cannot say if it will improve the simulation. 
Nevertheless, in \cref{Sec::Results} we present results detailing how this \gls{MPS}+\gls{MC} hybrid routine potentially reduces the severity of the sign problem.

\subsubsection*{Error sources}
Since standard \gls{AFQMC} would utilize \gls{HS}-decoupling to compute the trace exactly, this \gls{MPS} step introduces further imprecision to the method. 
Just as in any \gls{AFQMC}-approach, one source of error is splitting up non-commuting parts of the Hamiltonian on the order of $\mathcal{O}(\mathrm{d}\tau^2)$, i.e. the Trotter error.
Performing the \gls{HS} decoupling then adds in $\mathcal{O}(\mathrm{d}\tau^4)$ which is negligible in comparison. 
Additionally, as this is a \gls{QMC} algorithm there will also be statistical errors which can be obtained by a Jackknife estimate of the standard error.

More importantly, the newly added errors are those of the \gls{MPS} method. 
Primarily, this comes from limiting, by necessity, the bond dimension used to approximate the state. 
This yields an error to any observable computed using the \gls{MPS} which can be estimated via the truncation error~\cite{Schollwock2005,Schollwock2011}. 
Thus, any measurement taken from the algorithm now has an additional error, the magnitude of which we can easily estimate. 
In this manner, measurements will carry this additional error, and so too will partition functions.

Such imprecision has two effects: Measurement errors and incorrect update acceptance and rejection. 
The former simply occurs since each sample has the same \gls{MPS} error and averages to the same error when computing the measurement prediction. 
However, since the partition function also will have an error due to the \gls{MPS}-approximation, it can happen that acceptance or rejection is obtained erroneously. 
The probability of this occurring is on the same order as the partition functions imprecision. 
This is likely to have a lesser effect as the \gls{MPS} error has to be kept quite small. 
Nonetheless, the error propagation of this effect is unknown in the sense that we cannot know how often it occurs or what is the resulting error.

Finally, the \gls{MPS} numerics obtain errors when applying exponential operators to states. 
The size of this error depends heavily on the type of Hamiltonian and time evolution method chosen~\cite{Paeckel2019}. 
Fortuitously, this error is smaller than $\mathcal{O}(\mathrm{d}\tau^2)$ for most methods. 
Additionally, the splitting of Hamiltonian interaction terms performed in \cref{splitpartit} coincide with some evolution methods~(e.g.~TEBD). 
This makes the time evolution method error source of the same order as errors which already exist in the \gls{QMC} part of the algorithm.
\subsection{\label{Sub_Sec::Update_scheme}Update scheme}
Since part of this algorithm contains \gls{MC} sampling, the method will rely on proposing updates to an existing configuration and evaluating the updated trace. 
Acceptance of such an update is then governed by a Metropolis-Hastings algorithm.

While it is possible to simply evaluate the full \gls{1D} traces for each configuration, this turns out difficult to implement. 
Such a strategy could be considered global in the sense that every configuration field is updated at every iteration. 
This method has the benefit of changing all fields at once causing each sample to be weakly correlated to the next: All samples will be independent. 
Unfortunately, if we simply select a random configuration the probability of it being good enough for acceptance is intractably low.

One way to deal with this issue is to simply change a single configuration field and then evaluate the partition function. 
Such a strategy may choose a random value out of the three  choices for a given configuration field $l_{\tau,n}$, increasing the chance of acceptance. 
Yet \gls{MPS}-based algorithms will generally perform poorly in such an update scheme: they evolve an entire \gls{1D} system. 
While more local control is possible, in principle, such an algorithm would be challenging to implement for this proof-of-concept work. 
Instead, we may utilize the existing \gls{MPS} routines to simultaneously update all the \gls{HS}-variables in a single slice of discretized imaginary time. 
For such a strategy, relying on randomness to generate new updates is not possible: The number of field combinations to choose from scales like $4^{n_f\cdot L_x}$ where $n_f$ is the number of different \gls{HS} fields at a site, and $L_x$ the length of a chain. 
Thus, the exponential scaling will make even short chains difficult to update randomly.

In order to solve this issue some heuristic which allows the selection of probable configurations is necessary i.e. an update proposal algorithm. 
In general, unless we have complete knowledge of the system we cannot perform such proposition perfectly, but a suitable heuristic can improve the rejection ratio of the update algorithm.

In the case of \gls{Q1D} systems it is possible to utilize heuristics for proposing updates which assume $V_\perp\ll U$. 
Notably, the trace varies with configuration to a lesser degree if this condition is fulfilled, while the prefactor $\Gamma_\mathcal{C}$ has the same sizes regardless of $V_\perp$. 
Therefore, a routine which suggests updates for consideration based on $\Gamma_\mathcal{C}$ yields low rejection ratios for small enough $V_\perp$. 
Thus, one choice for the probability of considering an update is
\begin{equation}
	P_{\text{consider}}(\mathcal{C}\to\mathcal{C}') = \max\left(1,\frac{\Gamma_{\mathcal{C}'}}{\Gamma_\mathcal{C}}\right).
\end{equation}
%

In practice, as $V_\perp$ is increased the consideration heuristic will worsen: at a large enough value, the rejection ratio will become intractably large limiting the algorithm to small values of $V_\perp$. 
In principle, this property could be cured if an \gls{MPS} code which can efficiently update single sites at a time was available.

\section{\label{Sec::Results}Results}
In this section we present results on the performance of our \gls{MPS}-\gls{AFQMC} hybrid-algorithm for the two concrete models introduced in \cref{Sec::Model}. 
The primary focus lies in determining the severity of any sign problem compared to alternative methods. 
We divide the results into three subsections as follows: (\textit{A}) Studying the performance of the algorithm as compared with \gls{AFQMC}, (\textit{B}) analyzing the average sign of the hybrid algorithm and (\textit{C}) analyzing the effect of mass imbalance on the phase.

In the following results, all data has been generated with a time step size of $\mathrm{d}\tau=0.125$. 
The \gls{MPS}+\gls{MC} routine has kept the discarded weight at $\epsilon_\psi=10^{-8}$ resulting in an error at
\begin{equation}
	\epsilon_{\text{MPS}} \sim 10^{-4},
\end{equation}
which is the square root of $\epsilon_\psi$ as all measurements in this algorithm necessarily are brackets of two different many-body states~\cite{Schollwock2005}.

\subsection{\label{SubSec::Energy_and_SignProb}Energy and sign problem}
As our method aims to alleviate the sign problems for models of the type given in \cref{Eq::Hamiltonian}, it's important to assess how it compares to established algorithms. 
Therefore, we compute the internal energy of a system using both \gls{AFQMC} and our hybrid approach for the repulsive model defined by \cref{Eq::Hamiltonian} and \cref{Eq::Repulsive}.
\begin{figure}
	\ifthenelse{\boolean{buildtikzpics}}{
	\tikzsetnextfilename{MPSAFQMCComp}
	\tikzset{external/export next=true}
	\begin{tikzpicture}
		\begin{axis}
			[
				width = \columnwidth,
				xlabel = $\beta/t$,
				ymin = -1375,
				ymax = -1340,
			]
			\addplot+
			[
				colorA,
				only marks,
				mark=o,
				thick,
				error bars/.cd,
				y explicit,
				y dir = both,
			]
				table
				[
					y expr = \thisrowno{1},
					x expr = \thisrowno{0},
					y error expr = \thisrowno{2},
				]
			{Data/afqmc_ener_over_beta.dat};
			\addlegendentry{AFQMC}
			\addplot+
			[
				colorB,
				only marks,
				mark=o,
				thick,
				error bars/.cd,
				y explicit,
				y dir = both,
			]
				table
				[
					y expr = \thisrowno{1},
					x expr = \thisrowno{0},
					y error expr = \thisrowno{2},
				]
			{Data/mpsmc_ener_over_beta.dat};
			\addlegendentry{MPSMC}
		\end{axis}
	\end{tikzpicture}
	}{\includegraphics{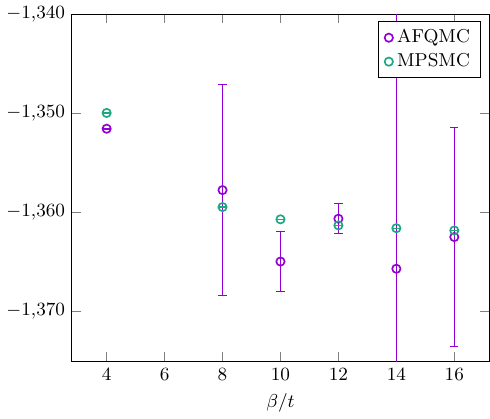}}
	\caption{Thermal energy plotted against imaginary time computed using the hybrid \gls{MPS}-\gls{QMC} algorithm and \gls{AFQMC} at parameters $L_y=L_x=16$, $t_\sigma=t$, $U=8t$, $t^\prime=0.25t$, $V_\perp=t$ and $\mu=-6.0$.}
	\label{Fig::AFQMC_MPSMC_comp}
\end{figure}
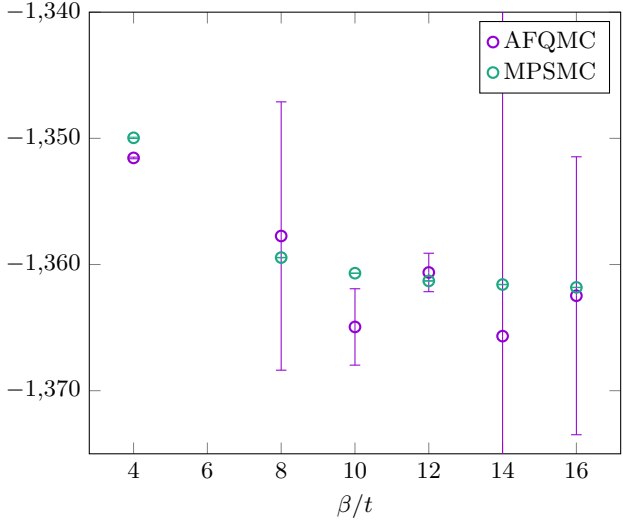
Rewriting the interaction term acting on every link $(i,j)-(i,j+1)$
%
\begin{align}\label{Eq::Op_split_model_1}
	\nonumber V_\perp\hat{W}_{i,j}\hat{W}_{i,j+1} & = \hat{V}_{(i,j)-(i,j+1)} + \frac{V_\perp}{2}\left(n_{i,j}+n_{i,j+1}\right) \\ & + V_\perp\left(n_{i,j,\uparrow}n_{i,j,\downarrow}+n_{i,j+1,\uparrow}n_{i,j+1,\downarrow}\right), \\
	\hat{V}_{(i,j)-(i,j+1)} & := -\frac{V_\perp}{2}\left(n^{\nodagger}_{i,j}-n^{\nodagger}_{i,j+1}\right)^2,
\end{align}
in line with~\cref{Eq::Newhamsplit,Eq::Splitidentity,Eq::Splitform}, shows that it can be brought to a form compliant with the algorithm described in \cref{Sec::Method}.
The price for this is the introduction of counter-terms, i.e. the last two terms on the right-hand side of~\cref{Eq::Op_split_model_1}.
Fortuitously, as these can be absorbed into $\sum_j \hat{H}^0_{1\text{D},j}$ (by the chemical potential- and on-site interaction-terms respectively) they are easily dealt with by the \gls{MPS}-component of our hybrid approach.

The comparison is shown in \cref{Fig::AFQMC_MPSMC_comp}.
It is important to note that studying a sign problem is partially a quantitative task: Given an amount of computing resources the algorithm which obtains the best estimate is preferred. 
As such, each point of $\beta$ data in \cref{Fig::AFQMC_MPSMC_comp} is computed using the same amount of core-hours for both algorithms.
With this restriction, it is clear that the exponential nature of the sign problem severely magnifies the error bar for \gls{AFQMC}.
In contrast, the hybrid algorithm our hybrid algorithm is free from any sign problem for this model, all configurations carrying the same sign.

Notably, the agreement between \gls{AFQMC} and the hybrid algorithm is not exact even when the sign problem is manageable.
This is expected due to restrictions imposed by the latter algorithm.
First, being an \gls{MPS} algorithm it benefits from utilizing quantum number symmetries. 
As such, we have computed the hybrid algorithm results in a quantum number sector where each \gls{1D} chain has zero overall spin. 
Conversely, \gls{AFQMC} keeps the ensemble grand canonical and will include states of various spin at finite temperature. 
Secondly, \gls{MPS} algorithms work significantly better with \gls{OBC} prompting us to use such conditions along the chains while using \gls{PBC} in the transverse direction. 
Not only does this affect the thermal state: it also removes a bond from the \gls{1D} system, raising the energy and causing the difference in internal energy as observed.

The missing sign problem for the hybrid algorithm begs the question: would a sign problem eventually appear at large enough $\beta$ or $L_x$ and $L_y$? 
In other words, it is uncertain from \cref{Fig::AFQMC_MPSMC_comp} whether the hybrid algorithm mitigates the sign problem or cures it completely.

%

\subsection{\label{SubSec::Results_SignProb}Sign problem in MPSMC}
\begin{figure}
	\ifthenelse{\boolean{buildtikzpics}}{
	\def\xmin{0.0}%
	\def\xmax{69.0}%
	\def\ymin{0.005}%
	\def\ymax{0.055}%
	\def\cbmin{0}%
	\def\cbmax{1}%
	\def\heatmapplotfilename{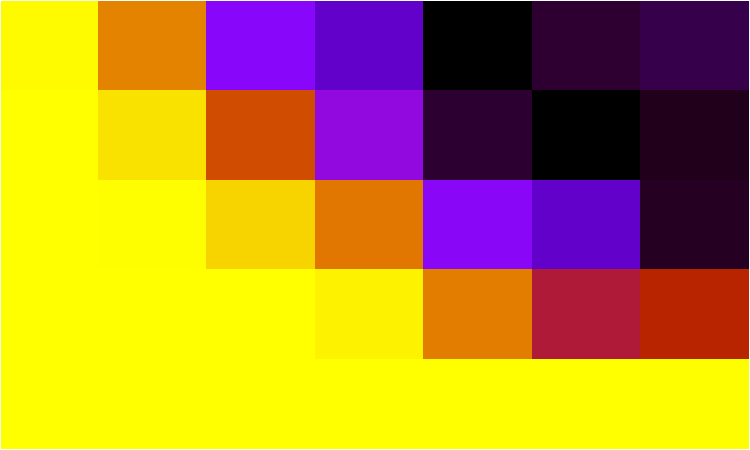}%
	\immediate\write18{echo "set terminal pdf; set output \string\"\heatmapplotfilename\string\"; unset border; unset xtics; unset ytics; unset colorbox; set lmargin 0; set rmargin 0; set tmargin 0; set bmargin 0; set cbrange[\cbmin:\cbmax]; p [\xmin:\xmax][\ymin:\ymax]\string\"Data/sign_hmap.dat\string\" u 1:2:3 w image" | gnuplot}%
	\pgfplotsset%
	{%
		/pgfplots/colormap={gnuplot}{rgb255=(0,0,0) rgb255=(46,0,53) rgb255=(65,0,103) rgb255=(80,0,149) rgb255=(93,0,189) rgb255=(104,1,220) rgb255=(114,2,242) rgb255=(123,3,253) rgb255=(131,4,253) rgb255=(139,6,242) rgb255=(147,9,220) rgb255=(154,12,189) rgb255=(161,16,149) rgb255=(167,20,103) rgb255=(174,25,53) rgb255=(180,31,0) rgb255=(186,38,0) rgb255=(191,46,0) rgb255=(197,55,0) rgb255=(202,64,0) rgb255=(208,75,0) rgb255=(213,87,0) rgb255=(218,100,0) rgb255=(223,114,0) rgb255=(228,130,0) rgb255=(232,147,0) rgb255=(237,165,0) rgb255=(241,185,0) rgb255=(246,207,0) rgb255=(250,230,0) rgb255=(255,255,0) }%
	}%
	\tikzsetnextfilename{SignHeatMap}
	\tikzset{external/export next=true}
	\begin{tikzpicture}%
		\begin{axis}%
			[%
			xmin			=	\xmin,%
			xmax			=	\xmax,%
			ymin			=	\ymin,%
			ymax			=	\ymax,%
			point meta min	=	\cbmin,%
			point meta max	=	\cbmax,%
			xlabel			=	{$\beta$},%
			ylabel			=	{$V_\perp$},%
			width			=	0.4\textwidth,%
			height			=	0.4\textwidth,%
			axis on top,
			every colorbar/.append style =%
			{%
				ylabel	=	{Average sign},%
				width	=	3mm,%
			},%
			colorbar,%
			axis on top,%
			]%
			\addplot graphics%
			[%
			xmin	=	\xmin,%
			xmax	=	\xmax,%
			ymin	=	\ymin,%
			ymax	=	\ymax,%
			]%
			{\heatmapplotfilename};%
		\end{axis}%
	\end{tikzpicture}%
	}{\includegraphics{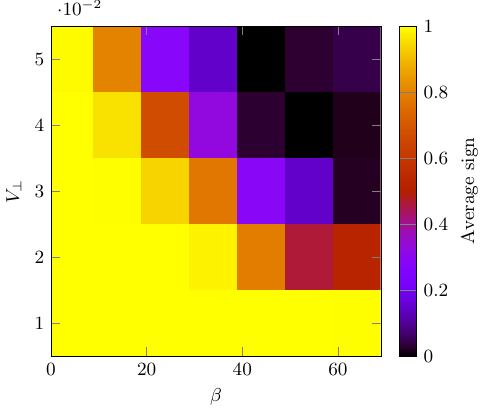}}
	\caption{Average sign for the effective hopping interaction model defined by \cref{Eq::EffHopPairCreat,Eq::EffHopPairAnn,Eq::EffHopUp,Eq::EffHopDown} for different $V_\perp$ and imaginary temperature $\beta$ at parameters $L_y=4$, $L_x=16$, $t_\uparrow=t$, $t_\downarrow=0.5t$, $t^\prime=0$, $U=-10t$, $\mu=5.09t$.}
	\label{Fig::MPSMC_sign_problem}
\end{figure}
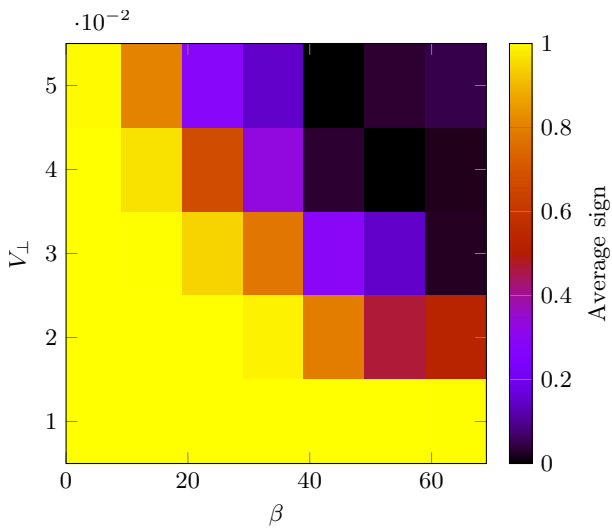

Since we have not found a regime in which the sign problem appears for the model of the previous section, we instead target a different one: the effective hopping interaction defined by \cref{Eq::Hamiltonian} and \cref{Eq::EffHopPairCreat,Eq::EffHopPairAnn,Eq::EffHopUp,Eq::EffHopDown}. 
As in~\cref{SubSec::Energy_and_SignProb}, this interaction is not directly amenable to \gls{HS} decoupling, as none of the transverse terms fulfill the decoupling condition, \cref{Eq::Splitform}. Neither is it immediately clear if there are counter-terms which can make an \gls{HS} decoupling possible.
Nevertheless, we present a set of counter-terms which can bring this interaction into a compliant form as well. 
For all the terms appearing in $\sum_j\hat{H}_{\perp,j}$ that connect link $(i,j)-(i,j+1)$ we can write
%
\begin{align}
	\nonumber V_\perp\sum_{k=1}^4 \hat{W}_{i,j}^{(k)} \hat{W}_{i,j+1}^{(k)\dagger} & = \sum_{k=1}^4 \hat{V}_{(i,j)-(i,j+1),k} \\
	\nonumber & + V_\perp( n_{i,j,\uparrow}n_{i,j,\downarrow}+n_{i,j+1,\uparrow}n_{i,j+1,\downarrow}) \\
	\label{Eq::Op_split_model_2} &- V_\perp(n_{i,j}+n_{i,j+1}) + \text{const.}
\end{align}
where we introduce
\begin{align}\label{Eq::PairInt}
	\nonumber \hat{V}_{(i,j)-(i,j+1),1} &:= \frac{V_\perp}{4}\Bigl[c^{\dagger}_{i,j,\uparrow}c^{\dagger}_{i,j,\downarrow}+c^{\dagger}_{i,j+1,\uparrow}c^{\dagger}_{i,j+1,\downarrow}\\&+c^{\nodagger}_{i,j,\downarrow}c^{\nodagger}_{i,j,\uparrow}+c^{\nodagger}_{i,j+1,\downarrow}c^{\nodagger}_{i,j+1,\uparrow}\Bigr]^2, \\
	\label{Eq::ImagPairInt}\nonumber \hat{V}_{(i,j)-(i,j+1),2} &:= \frac{V_\perp}{4}\Bigl[i\Bigl(c^{\dagger}_{i,j,\uparrow}c^{\dagger}_{i,j,\downarrow}+c^{\dagger}_{i,j+1,\uparrow}c^{\dagger}_{i,j+1,\downarrow}\\&-(c^{\nodagger}_{i,j,\downarrow}c^{\nodagger}_{i,j,\uparrow}+c^{\nodagger}_{i,j+1,\downarrow}c^{\nodagger}_{i,j+1,\uparrow})\Bigr)\Bigr]^2, \\
	\label{Eq::SpinUpInt}\hat{V}_{(i,j)-(i,j+1),3} &:= -\frac{V_\perp}{2}\left(n_{i,j,\uparrow}-n_{i,j+1,\uparrow}\right)^2, \\
	\label{Eq::SpinDownInt}\hat{V}_{(i,j)-(i,j+1),4} &:= -\frac{V_\perp}{2}\left(n_{i,j,\downarrow}-n_{i,j+1,\downarrow}\right)^2.
\end{align}
As for the previous model, this rewriting only yields counter-terms that are easily absorbed into the \gls{1D} part of the Hamiltonian, and thus easily dealt with by our approach's \gls{MPS}-component.
Thus, using the new interaction definitions \cref{Eq::PairInt,Eq::ImagPairInt,Eq::SpinUpInt,Eq::SpinDownInt} we obtain four terms which all may be \gls{HS} decoupled using \cref{gausshermite}. 
Computing the average sign of all configurations for different $V_\perp$ and imaginary time $\beta$ sees the sign problem appearing at low temperatures and stronger coupling as shown in \cref{Fig::MPSMC_sign_problem}.
Notably, even a weak $V_\perp$ yields a quickly decaying average sign at somewhat modest $\beta$ concluding that the hybrid algorithm does not generally cure the sign problem.

\subsection{Mass imbalance phase}
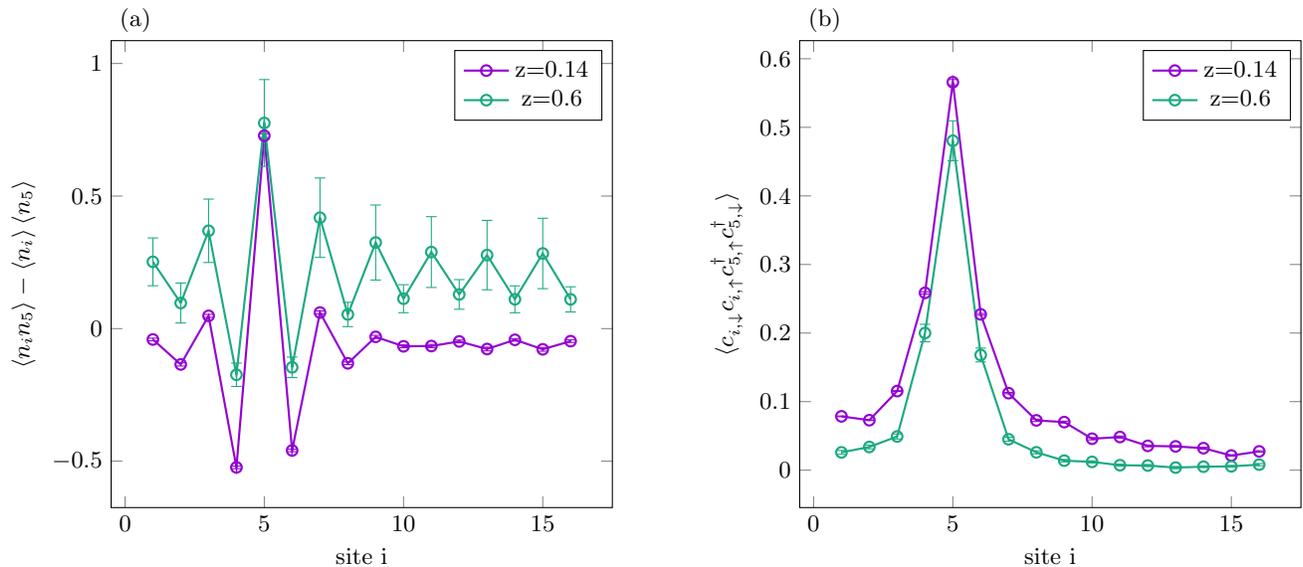
\begin{figure*}[t!]
	\ifthenelse{\boolean{buildtikzpics}}
	{
	\tikzsetnextfilename{MassImbPhaseDensCorr}
	\tikzset{external/export next=true}
	\begin{tikzpicture}
		\begin{axis}
		[
			width=0.46\textwidth,
			height=0.33\textheight,
			xlabel = site i,
			ylabel=$\braket{n_in_5}-\braket{n_i}\braket{n_5}$,
			title = (a),
			every axis title/.style={above right, at={(0,1)}},
		]
		\addplot+
		[
			colorA,
			mark=o,
			thick,
			error bars/.cd,
			y explicit,
			y dir = both,
		]
		table
		[
			y expr = \thisrow{nncorr},
			x expr = \thisrow{site},
			y error expr = \thisrow{err},
		]
		{Data/nn_corr_z_0.14.dat};
		\addlegendentry{z=0.14}
		\addplot+
		[
			colorB,
			mark=o,
			thick,
			error bars/.cd,
			y explicit,
			y dir = both,
		]
		table
		[
		y expr = \thisrow{nncorr},
		x expr = \thisrow{site},
		y error expr = \thisrow{err},
		]
		{Data/nn_corr_z_0.6.dat};
		\addlegendentry{z=0.6}
		\end{axis}
		\end{tikzpicture}
		}{\includegraphics{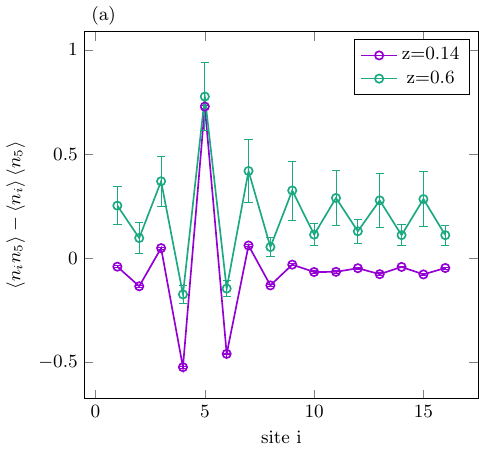}}
		\hspace{3em}
		\ifthenelse{\boolean{buildtikzpics}}{
		\tikzsetnextfilename{MassImbPhasePairCorr}
		\tikzset{external/export next=true}
		\begin{tikzpicture}
		\begin{axis}
		[
			width=0.46\textwidth,
			height=0.33\textheight,
			xlabel = site i,	ylabel=$\braket{c^\nodagger_{i,\downarrow}c^\nodagger_{i,\uparrow}c^\dagger_{5,\uparrow}c^\dagger_{5,\downarrow}}$,
			title = (b),
			every axis title/.style={above right, at={(0,1)}},
		]
		\addplot+
		[
			colorA,
			mark=o,
			thick,
			error bars/.cd,
			y explicit,
			y dir = both,
		]
			table
			[
				y expr = \thisrow{pcorr},
				x expr = \thisrow{site},
				y error expr = \thisrow{err},
			]
		{Data/pair_corr_z_0.14.dat};
		\addlegendentry{z=0.14}
		\addplot+
		[
			colorB,
			mark=o,
			thick,
			error bars/.cd,
			y explicit,
			y dir = both,
		]
			table
			[
				y expr = \thisrow{pcorr},
				x expr = \thisrow{site},
				y error expr = \thisrow{err},
			]
		{Data/pair_corr_z_0.6.dat};
		\addlegendentry{z=0.6}
	\end{axis}
	\end{tikzpicture}
	}{\includegraphics{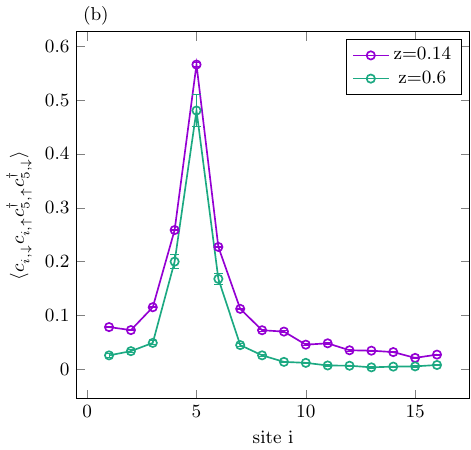}}
	\caption{A comparison of correlators measured along a single chain at equal density for mass imbalance $z=0.14$ with $\mu=5.0475t$ and $z=0.6$ with $\mu=5.04t$ at $t^\prime=0$, $U=-10t$, $\beta=64.0t$, $V_\perp=0.01t$, $L_x=16$ and $L_y=4$. Sub-figure (a) shows the disconnected density-density correlator and (b) the pair-pair correlator.}
	\label{Fig::MassImbalance}
\end{figure*}

We turn to a study of two different mass imbalance parameters system away from half filling. 
Such systems have been studied previously via analytical methods~\cite{Cazalilla2005}. 
From this previous work, attractive on-site interaction favors a superconducting phase at small imbalances while shifting to a \gls{CDW} phase at larger values. 
We parametrize the imbalance using
\begin{equation}
	z = \frac{|t_\uparrow -t_\downarrow|}{|t_\uparrow+t_\downarrow|},
\end{equation}
and take correlator measurements at two different $z$-values, as presented in \cref{Fig::MassImbalance}.
We have adjusted the chemical potential to obtain equal density at both values of $z$ for suitable comparison which results in a density of $n=0.890625$.
Notably, the pairing correlator decays quickly for both values of the imbalance parameter $z$ but does not manage to reach zero for this system size as shown in \cref{Fig::MassImbalance}b.
Conversely, the density-density correlator decays quickly to zero for smaller values of imbalance. 
When imbalance is increased to moderate values there is no longer an apparent decay to zero and a strong oscillating behavior is obtained.
We thus obtain the first quantitative results confirming the qualitative analytical predictions that the growth of $z$ drives even systems that are doped away from half filling away from a superconducting and towards a \gls{CDW} instability.
The proof-of-concept shown here thus makes it possible for future work to systematically quantify this effect and calculate a phase diagram.
 
\section{\label{Sec::Discussion}Discussion}
We have developed an algorithm, containing only controlled approximations, which combines \gls{MPS}-based numerical methods and \gls{AFQMC} in order to solve models of \gls{Q1D} quantum many-body systems from first principles. 
For repulsive interactions at arbitrary density (i.e. away from unit filling) such systems cannot generally be solved at the required sizes and/or temperatures using either one of these methods alone: \gls{MPS} numerics suffers from the exponentially growing entanglement in \gls{2D} and \gls{3D} systems, while \gls{AFQMC} usually has a sign problem for such systems. 
We find that our new algorithm outperforms \gls{AFQMC} in these models for equal amounts of resources when studying repulsive Hubbard chains with triangular lattice structure connected via repulsive interaction. 
The primary reason for the observed advantage is that we do not find any sign problem using the new algorithm.

Encouraged by this result we attempt to find whether a sign problem may still be obtained by the hybrid algorithm. 
Studying a mass-imbalanced system with attractive interactions away from half filling, which is connected by both repulsive interaction and Josephson coupling, we vary the transverse connection strength, $V_\perp$. 
This clearly obtains a sign problem indicating that the method does not completely cure the sign problem, but alleviates it. 

One potential explanation for this is that the added Josephson coupling introduces complex terms to the \gls{1D} Hamiltonian when decoupled. 
However, it is important to keep in mind that it may well be a combination of multiple issues which generates the problem, e.g. that we now have two competing orders for the system to choose from.

Nevertheless, we employ our new algorithm to study the shift from a superconducting to a \gls{CDW} instability for this sign-problem-laden system. 
This transition has been studied with qualitative field-theory methods previously. 
In this paper it is performed away from half filling, yielding results which would typically be difficult to obtain using alternative algorithms. 
Limited by the sign problem in what system size we may simulate, we find the transition occurring between small and moderate values of mass imbalance, consistent with the bonsonization results.

\section{Outlook\label{Sec::Outlook}}
This work presents a proof-of-concept analysis of a novel algorithm. 
Comprehensive determination as to which systems may benefit from its application requires further work. 
But we already demonstrate here that our algorithm addresses the sign problem to different degrees for different models.
We reduce or eliminate the sign problem that an \gls{AFQMC}-only treatment would incur as we offload those elements that are known to cause it - local repulsion in doped systems, frustration - to the \gls{MPS}-component of the algorithm.
Conversely, the parts of the Hamiltonian that lead to the system being higher-dimensional than \gls{1D} are handled by the \gls{AFQMC}-component of our hybrid approach, which strongly reduces the amount of entanglement the \gls{MPS}-component needs to retain.

Subsequent research should focus on the detailed mechanisms by which the two component algorithms achieve these useful effects. 
Specifically, since there is no sign problem in Model~1, in which there are only repulsive interactions in the $y$-direction, one line for further inquiry should be whether the sign problem in Model~2 appears as a consequence of the two competing instability-channels of that model. 
Should this turn out to be the case, an immediate remedy for the sign-problem would be to perform simulations only one channel at a time and compare free energies in order to determine which phase is dominant.

Regarding the currently-used update scheme of \cref{Sub_Sec::Update_scheme}, and its practical limitation to parameter regimes of weak coupling constant $V_\perp$: this limitation can be overcome by development of a custom \gls{MPS} algorithm capable of performing efficient single-site Metropolis-Hastings updates.
Such a routine could commence from testing the performance of an approach that efficiently evaluates the expectation values of differential update operators $e^{\sqrt{\tau}\Delta\eta_{\tau,n}a_i^{(n)} \hat{o}_i}$.
Here, $\Delta\eta_{\tau,n}$ denotes the difference between the current value of the auxiliary field $\eta_{\tau,n}$ and the proposed new value.

Finally, this work has focused on the statistical error stemming from the \gls{MC} part of the algorithm. 
As noted in \cref{Sec::Results}, there are additional error sources which are of smaller magnitude. 
Future work would benefit from analyzing how robust the output is under varying \gls{MPS} errors, i.e. bond dimension.

The hybrid algorithm proposed here could be useful to a number of systems where obtaining the equilibrium properties at size and low temperatures is a significant current challenge.
Systems as diverse as Rydberg atom arrays, ultracold dipolar gases as well as neutral atoms confined to optical cavities, trapped ion setups and practically any effective Hamiltonian in quantum chemistry feature long-ranged interactions of the type that would be amenable to the \gls{HS}-decoupling at the root of our hybrid algorithm~\cite{Al-Saidi2006,Purwanto2009,Defenu2023}.
These system could thus be split into several clusters, for which the larger shorter-range couplings are handled by the \gls{MPS}-component of the algorithm, thus sidestepping any sign-problem.
Longer-ranger couplings in-between clusters would then be handled by the AFQMC-component of the algorithm.
Another interesting translation of the approach could be to the recently described superconducting nickelates La$_3$Ni$_2$O$_7$, which feature inter-plane coupling that is purely of a spin-spin type~\cite{Oh2023,Qu2023}, and thus amenable to \gls{HS}-decoupling.
Analogous to the present work, the in-plane physics could be treated with a generalized tensor-network approach such as PEPS, while the inter-plane physics is handled by the AFQMC-component.
%
%
\begin{acknowledgments}
This work was supported by an ERC Starting Grant from the European Union's Horizon 2020 research and innovation programme under grant agreement No. 758935; and the UK's Engineering and Physical Sciences Research Council [EPSRC; grant number EP/W022982/1]. 
The computations were enabled by resources provided through multiple EPSRC ``Access to HPC" calls (Spring 2023, Autumn 2023 and Spring 2024) on the ARCHER2, Peta4-Skylake and Cirrus compute clusters, as well as by compute time awarded by the National Academic Infrastructure for Supercomputing in Sweden (NAISS).
\end{acknowledgments}

\bibliography{gunnar,adrian}
\end{document}

%% file: header.tex
\usepackage{lineno}
\usepackage{graphicx}  
\usepackage{dcolumn}   
\usepackage{bm}        
\usepackage{amssymb}   
\usepackage{amsmath}
\usepackage{blkarray, multirow, graphicx, diagbox, color, xcolor, colortbl}
\usepackage{bbm, bbold}
\usepackage{glossaries}
\usepackage{hyphenat}
\usepackage{ifthen}
\usepackage{xkeyval}
\usepackage{moreverb}
\usepackage{rotating}
\usepackage{wrapfig}
\usepackage{slashbox}
\usepackage{xspace}
\usepackage{nicefrac}
\usepackage[]{units}
\usepackage{physics}
\usepackage{braket}
\usepackage[inline]{enumitem}
\usepackage{tabto}
\usepackage{listings}
\usepackage{xstring}
\def\ReplaceStr#1{%
	\IfSubStr{#1}{p}{%
		\StrSubstitute{#1}{p}{.}}{#1}}

\usepackage[caption=false]{subfig}
\usepackage{pgf}
\usepackage{tikz}
\tikzset{>=stealth}
\usepackage{calc}
\usepackage{pgffor}
\usepackage{pgfplots}
\pgfplotsset{compat=newest}
\usepackage{pgfplotstable}
\usepgfplotslibrary{groupplots}

\tikzstyle{n} = [draw,shape=ellipse,minimum size=1.5em,inner sep=0pt,fill=white!20, minimum width=2.5em]
\tikzstyle{Init} = [n,color=green,fill=green!20,text=black]
\tikzstyle{Fin} = [n,color=red,fill=red!20,text=black]
\tikzstyle{Ghost} = [minimum size=1.5em,inner sep=0pt,color=white,text=black]
\tikzstyle{Multiple} = [draw,shape=rect,minimum size=2em,inner sep=0pt]

\tikzstyle{ghostA} = [text=red!70,thick, minimum size=2*(5pt-\pgflinewidth), inner sep=0pt, outer sep=0pt]
\tikzstyle{ghostB} = [text=blue!70,thick, minimum size=2*(5pt-\pgflinewidth), inner sep=0pt, outer sep=0pt]
\tikzstyle{siteA} = [regular polygon, regular polygon sides=3, shape border rotate= 30, draw=red!50,fill=red!20,thick,inner sep=0pt,minimum width=1.5em,font=\footnotesize]
\tikzstyle{siteB} = [regular polygon, regular polygon sides=3, shape border rotate= -30, draw=green!50,fill=green!20,thick,inner sep=0pt,minimum width=1.5em,font=\footnotesize]
\tikzstyle{op} = [regular polygon, regular polygon sides=4, draw=orange!50, fill=orange!20, thick, inner sep=0.2pt, minimum width=1.25em, minimum height=1.5em,font=\footnotesize]
\tikzstyle{opghost} = [regular polygon, regular polygon sides=4, thick, inner sep=0.2pt, minimum width=1.25em, minimum height=1.5em,font=\footnotesize]
\tikzstyle{site} = [circle,draw=blue!50,fill=blue!20,thick,inner sep=0.2pt,minimum width=1.25em,font=\footnotesize]
\tikzstyle{hiddensite} = [circle,draw=white!50,fill=white!20,thick,inner sep=0.2pt,minimum width=1.25em,font=\footnotesize]
\tikzstyle{nosite} = [circle,draw=white,fill=white,thick,inner sep=0.1pt,minimum width=1.5em]
\tikzstyle{ghost} = [font=\footnotesize]
\tikzstyle{intersite} = [regular polygon, regular polygon sides=4, shape border rotate= 45, draw=black!50,fill=black!20,thick,inner sep=0pt,minimum width=1.5em]
\tikzstyle{ld} = [inner sep=1pt, font=\small]
\tikzstyle{unsite} = [circle, outer sep=0pt,inner sep=0.2pt,minimum width=1.25em]

\definecolor{colorA}{rgb} {0.58,0,0.8275}
\definecolor{colorB}{rgb} {0.11,0.663,0.51}
\definecolor{colorC}{rgb} {0.3373,0.7059,0.9137}
\definecolor{colorD}{rgb} {0.902,0.6235,0}
\definecolor{colorE}{rgb} {0.9451,0.902,0.3255}
\definecolor{colorF}{rgb} {0.3373,0.3255,0.902}
\definecolor{colorG}{rgb} {0.9451,0.3255,0.3373}

\usetikzlibrary
{
	calc,
	decorations,
	calligraphy,
	pgfplots.fillbetween,
	pgfplots.patchplots,
	plotmarks,
	patterns,
	positioning,
	petri,
	arrows,
	intersections,
	decorations.markings,
	backgrounds,
	fit,
	matrix,
	graphs,
	shapes.geometric,
	decorations.pathreplacing, 
	decorations.pathmorphing,
	decorations,
	shapes.misc,
	shapes.multipart,
	shapes,
	through,
	tikzmark,
	pgfplots.colorbrewer,
	fpu,
}

\pgfplotsset{
        cycle from colormap manual style/.style={
            x=3cm,y=10pt,ytick=\empty,
            stack plots=y,
            every axis plot/.style={line width=2pt},
        },
}

\tikzset{->-/.style={decoration={
			markings,
			mark=at position .5 with {\arrow{>}}},postaction={decorate}}}

\tikzset{-<-/.style={decoration={
			markings,
			mark=at position .5 with {\arrow{<}}},postaction={decorate}}}

\tikzstyle{orientedsnake} = [
decorate, 
decoration={snake},
->
]  
\tikzstyle{orientedshortarrow} = [
decoration={markings,
	mark=at position .33 with {\arrow{>}}},
postaction={decorate}
]  
\tikzstyle{orientedlongarrow} = [
decoration={markings,
	mark=at position .67 with {\arrow{>}}},
postaction={decorate}
]
\tikzset{dbl/.style={double,
		double equal sign distance,
		-implies,
		shorten >=10pt,
		shorten <=10pt}}
\tikzset{
	between/.style args={#1 and #2}{
		at = ($(#1)!0.5!(#2)$)
	}
}

\newcommand{\nodagger}[0]{{\phantom{\dagger}}}

\pgfmathdeclarefunction{peierlspotential}{6}%
{
	\pgfmathparse
	{
		#2 / (#3 * #5) 
		* sin(deg(#1 * #3 - #3 * #5 * (x)))
		* exp(-( ( #1 - #5 * ((x) - #4) ) )^2.0 / ( #6^2.0 ) )
	}%
}

\pgfmathdeclarefunction{linearFct}{2}%
{%
	\pgfmathparse{#1*x+#2}%
}
\pgfmathdeclarefunction{quadFct}{3}%
{%
	\pgfmathparse{#1*x*x+#2*x+#3}%
}
\pgfmathdeclarefunction{logFct}{2}%
{%
	\pgfmathparse{#1*log10(x)+#2}%
}
\pgfmathdeclarefunction{expFct}{2}%
{%
	\pgfmathparse{#1*exp(-x/#2)}%
}

\newboolean{buildCurrentBlockInline}
\setboolean{buildCurrentBlockInline}{false}
\newboolean{rebuildCurrentBlock}
\setboolean{rebuildCurrentBlock}{false}

\newboolean{rebuildData}
\setboolean{rebuildData}{false}

\newboolean{removeData}
\setboolean{removeData}{false}

\newboolean{buildtikzpics}
\setboolean{buildtikzpics}{false}
\newif\ifrebuildtikz
\newif\ifChangeMode
\ChangeModetrue
\ChangeModefalse
\ifthenelse{\boolean{buildtikzpics}}
{
	\rebuildtikztrue
	\usetikzlibrary{external}
	\tikzexternalize[optimize=false,prefix=Figures/autogen/]
}
{
	\rebuildtikzfalse
}
\usepackage[colorlinks,bookmarks=false,citecolor=blue,linkcolor=black,urlcolor=blue]{hyperref}
\usepackage{cleveref}
\Crefname{appendix}{Appendix}{Appendices}
\Crefname{equation}{Equation}{Equations}
\Crefname{figure}{Figure}{Figures}
\Crefname{section}{Section}{Sections}
\Crefname{tabular}{Tabular}{Tabulars}
\crefname{appendix}{App.}{Apps.}
\crefname{equation}{Eq.}{Eqs.}
\crefname{figure}{Fig.}{Figs.}
\crefname{section}{Sec.}{Secs.}
\crefname{tabular}{Tab.}{Tabs.}

\ExplSyntaxOn
\DeclareExpandableDocumentCommand \eval { m } { \fp_eval:n { #1 } }
\ExplSyntaxOff

\makeatletter
\def\pgfplotsutil@decstringcounter#1{%
 \begingroup
  \c@pgf@counta=#1\relax
  \advance\c@pgf@counta by -1
  \edef#1{\the\c@pgf@counta}%
  \pgfmath@smuggleone#1%
 \endgroup
}%

\pgfplotsset{
/pgfplots/each nth point*/.style 2 args={%
/pgfplots/x filter/.append code={%
 \ifnum\coordindex=0
  \def\c@pgfplots@eachnthpoint@xfilter{0}%
  \edef\c@pgfplots@eachnthpoint@xfilter@cmp{#1}%
 \else
  \ifnum\coordindex>#2\relax
   \pgfplotsutil@advancestringcounter\c@pgfplots@eachnthpoint@xfilter
   \ifx\c@pgfplots@eachnthpoint@xfilter@cmp\c@pgfplots@eachnthpoint@xfilter
    \def\c@pgfplots@eachnthpoint@xfilter{0}%
   \else
    \let\pgfmathresult\pgfutil@empty
   \fi
  \fi
 \fi
}%
},
}
\makeatother

%% file: acronyms.tex
\newacronym[shortplural={MPS}]{MPS}{MPS}{matrix\hyp product state}
\newacronym{MF}{MF}{mean\hyp field}
\newacronym{BCS}{BCS}{Bardeen\hyp Cooper\hyp Schrieffer}
\newacronym{ACA}{ACA}{adaptive cross approximation}
\newacronym{1D}{1D}{one\hyp dimensional}
\newacronym{2D}{2D}{two\hyp dimensional}
\newacronym{3D}{3D}{three\hyp dimensional}
\newacronym{MPO}{MPO}{matrix\hyp product operator}
\newacronym{SVD}{SVD}{singular\hyp value decomposition}
\newacronym{QCS}{QCS}{quantum\hyp computer simulator}
\newacronym{QC}{QC}{quantum computer}
\newacronym{FSM}{FSM}{finite\hyp state machine}
\newacronym{CDW}{CDW}{charge\hyp density wave}
\newacronym{SDW}{SDW}{spin\hyp density wave}
\newacronym{ARPES}{ARPES}{angle-resolved photoemission spectroscopy}
\newacronym{OBC}{OBC}{open-boundary conditions}
\newacronym{PBC}{PBC}{periodic-boundary conditions}
\newacronym{TEBD}{TEBD}{time-evolution block-decimation}
\newacronym{TDVP}{TDVP}{time\hyp dependent variational principle}
\newacronym{iff}{iff}{if and only if}
\newacronym{DFT}{DFT}{density\hyp functional theory}
\newacronym{DMFT}{DMFT}{dynamical mean\hyp field theory}
\newacronym{DMRG}{DMRG}{density\hyp matrix renormalization group}
\newacronym{1DMRG}{1DMRG}{single-site density\hyp matrix renormalization group}
\newacronym{2DMRG}{2DMRG}{two-site density\hyp matrix renormalization group}
\newacronym{DMRG3S}{DMRG3S}{strictly single-site density\hyp matrix renormalization group}
\newacronym{iDMRG}{iDMRG}{inifinite\hyp size density\hyp matrix renormalization group}
\newacronym{tDMRG}{tDMRG}{time\hyp dependent density\hyp matrix renormalization group}
\newacronym{PP-DMRG}{PP-DMRG}{projected purified density\hyp matrix renormalization group}
\newacronym{QMC}{QMC}{quantum Monte Carlo}
\newacronym{MC}{MC}{Monte Carlo}
\newacronym{AIM}{AIM}{Anderson impurity model}
\newacronym{SIAM}{SIAM}{single impurity Anderson model}
\newacronym{LDA}{LDA}{local\hyp density approximation}
\newacronym{VQE}{VQE}{variational\hyp quantum eigensolver}
\newacronym{ED}{ED}{exact diagonalization}
\newacronym{QPT}{QPT}{quantum phase transition}
\newacronym{QCP}{QCP}{quantum critical point}
\newacronym{ETH}{ETH}{eigenstate thermalization hypothesis}
\newacronym{EHM}{EHM}{extended Hubbard model}
\newacronym{AKLT}{AKLT}{Affleck\hyp Lieb\hyp Kennedy\hyp Tasaki}
\newglossaryentry{QR}{name={QR},description={QR decomposition}}
\newacronym{TNS}{TNS}{tensor\hyp network state}
\newacronym{SM}{SM}{supplemental material}
\newacronym{NOO}{NOO}{natural orbital occupation}
\newacronym{NO}{NO}{natural orbital}
\newacronym{LRO}{LRO}{long\hyp range order}
\newacronym{qLRO}{qLRO}{quasi\hyp long\hyp range order}
\newacronym{SC}{SC}{superconductivity}
\newacronym{VBGS}{VBGS}{valence bond ground-state}
\newacronym{PEPS}{PEPS}{projected entangled pair\hyp states}
\newacronym{ALS}{ALS}{alternating least squares}
\newacronym{BdG}{BdG}{Bogoljubov de-Gennes}
\newacronym{TFIM}{TFI}{transverse field Ising model}
\newacronym{PP}{PP}{projected purification}
\newacronym{BEC}{BEC}{Bose\hyp Einstein condensate}
\newacronym{JWT}{JWT}{Jordan\hyp Wigner transformation}
\newacronym{NISQ}{NISQ}{noisy intermediate scale quantum}
\newacronym{NN}{NN}{nearest\hyp neighbor}
\newacronym{NNN}{NNN}{next\hyp nearest\hyp neighbor}
\newacronym{SPDM}{SPDM}{single\hyp particle density matrix} 
\newacronym{HCB}{HCB}{hardcore bosons}
\newacronym{SF}{SF}{spinless fermions}
\newacronym{fRG}{fRG}{functional renormalization group}
\newacronym{LE}{LE}{Luther\hyp Emery}
\newacronym{TLL}{TLL}{Tomonaga-Luttinger liquid}
\newacronym{PLL}{PLL}{pair Luttinger liquid}
\newacronym{FB}{FB}{flat band}
\newacronym{SU}{SU}{superfluid}
\newacronym{QO}{QO}{quasi-order}
\newacronym{LL}{LL}{Luttinger liquid}
\newacronym{CFT}{CFT}{conformal field theory}
\newacronym{USC}{USC}{unconventional superconductivity}
\newacronym{Q1D}{Q1D}{quasi-one-dimensional}
\newacronym{AFQMC}{AFQMC}{auxiliary-field quantum Monte Carlo}
\newacronym{HS}{HS}{Hubbard-Stratonovich}

%% file: tikzdefs.tex
\newcommand{\Systems}[7]%
{%
	\pgfmathparse{int((#2*#5-1))}%
	\pgfmathsetmacro{\ymax}{\pgfmathresult}%
	\pgfmathparse{int(#1-1)}%
	\pgfmathsetmacro{\xmax}{\pgfmathresult}%
	\tikzset{external/export next=false}%
	\begin{tikzpicture}[baseline]%
		\foreach \x in {0,...,\xmax}%
		{%
			\foreach \y in {0,...,\ymax}%
			{%
				\node[circle, draw, inner sep = 0pt, outer sep = 0pt] at ($(\x*1.5*#3,\y*1.25*#3)$) (\x\y){\pgfmathparse{int(10*#3)}\fontsize{\pgfmathresult pt}{10pt}#4};%
			}%
		}%
		\foreach \x [remember=\x as \lastx (initially 0)] in {0,...,\xmax}%
		{%
			\foreach \y [remember=\y as \lasty (initially 1)] in {0,...,\ymax}%
			{%
				\ifthenelse{\y>0}%
				{%
					\pgfmathparse{int(mod(\y,#2))}%
					\ifthenelse{\pgfmathresult=0}%
					{%
						\draw[densely dotted] (\x\lasty) -- node[left, inner ysep = 0.5em] {} (\x\y);%
					}%
					{%
						\draw (\x\lasty) -- node[left, inner ysep = 0.5em] {} (\x\y);%
					}%
				}{}%
				\ifthenelse{\ymax>0}%
				{%
					\draw[densely dotted, opacity=#7] (\lastx\lasty) -- node[midway] (m) {} (\x\y);%
				}%
				{}%
				\ifthenelse{\x>0}%
				{%
					\pgfmathparse{int(mod(\x,#2))}%
					\ifthenelse{\pgfmathresult=0}%
					{%
						\begin{scope}[on background layer]
							\draw[opacity=#6] (\lastx\y.east) -- node[above, inner xsep = 0.5em, inner ysep = 0.1em] {} (\x\y.west);%
						\end{scope}
					}%
					{%
						\draw (\lastx\y) -- node[above, inner xsep = 0.5em, inner ysep = 0.1em] {} (\x\y);%
					}%
				}{}%
			}%
		}%
		\pgfmathparse{int((#5-1)/2*(#2))}%
		\pgfmathsetmacro{\yPhysMin}{\pgfmathresult}%
		\pgfmathparse{int(\yPhysMin+#2-1)}%
		\pgfmathsetmacro{\yPhysMax}{\pgfmathresult}%
		\pgfmathparse{int(#1-1)}%
		\pgfmathsetmacro{\xmax}{\pgfmathresult}%
		\node[draw, fit = (0\yPhysMin) (\xmax\yPhysMax)] (Phys) {};
		\node[anchor=west] at (Phys.north east) {$H_0$};
	\end{tikzpicture}%
}%
\newcommand{\Fermions}[5]%
{%
	\pgfmathparse{rnd}%
	\pgfmathsetmacro{\uprnd}{\pgfmathresult}%
	\ifthenelse{\lengthtest{\uprnd pt > #1pt }}%
	{%
		{\color{#2}$\uparrow$}%
	}%
	{%
		\color{#3}$\uparrow$%
	}%
	\pgfmathparse{rnd}%
	\pgfmathsetmacro{\downrnd}{\pgfmathresult}%
	\ifthenelse{\lengthtest{\downrnd pt > #1pt }}%
	{%
		{\color{#4}$\downarrow$}%
	}%
	{%
		{\color{#5}$\downarrow$}%
	}%
}%
\newcommand{\Bosons}[3]%
{%
	\pgfmathparse{rnd}%
	\pgfmathsetmacro{\rnd}{\pgfmathresult}%
	\ifthenelse{\lengthtest{\rnd pt > #1pt }}%
	{%
		{\color{#2}\textbullet}%
	}%
	{%
		{\color{#3}\textbullet}%
	}%
}%

\definecolor{colorA}{rgb} {0.58,0,0.8275}
\definecolor{colorB}{rgb} {0.11,0.663,0.51}
\definecolor{colorC}{rgb} {0.3373,0.7059,0.9137}
\definecolor{colorD}{rgb} {0.902,0.6235,0}
\definecolor{colorE}{rgb} {0.9451,0.902,0.3255}
\definecolor{colorF}{rgb} {0.3373,0.3255,0.902}
\definecolor{colorG}{rgb} {0.9451,0.3255,0.3373}
\definecolor{colorH}{rgb} {0.11,0.3255,0.3373}

\pgfmathdeclarefunction{critExponentFct}{3}%
{%
	\pgfmathparse{(x<#2)?(#1*abs(x-#2)^(#3)):(0)}%
}

\pgfmathdeclarefunction{exponentialInFct}{3}%
{%
	\pgfmathparse{#1+#2*exp(-(#3)/x)}%
}

\newcommand{\printpgfnumberwithouterrorInMath}[3][0]%
{%
	\pgfmathfloatparsenumber{#2}%
	\pgfmathfloattomacro{\pgfmathresult}{\Fn}{\Mn}{\En}%
	\pgfmathparse{#1==2 ? (\Fn==2 ? "+" : "-") : (\Fn==2 ? "-" : (#1==1 ? "+" : ""))}%
	\edef\Sn{\pgfmathresult}%
	\pgfmathfloatparsenumber{#3}%
	\pgfmathfloattomacro{\pgfmathresult}{\Fe}{\Me}{\Ee}%
	\pgfmathparse{int(sqrt((\Ee-\En)^2))}%
	\edef\precisionAbsEe{\pgfmathresult}%
	\pgfmathparse{int(\Ee-\En)}%
	\edef\precisionE{\pgfmathresult}%
	\pgfmathparse{\eval{\Me*10^(\precisionE)}}%
	\ifthenelse{\En=0}%
	{%
		\Sn\pgfmathprintnumber[fixed, precision=\precisionAbsEe, zerofill]{\Mn}%
	}%
	{%
		\ifthenelse{\En=1}%
		{%
			\pgfmathparse{\eval{\Mn*10}}%
			\edef\Mn{\pgfmathresult}%
			\pgfmathparse{int(\precisionAbsEe+1)}%
			\edef\precisionAbsEe{\pgfmathresult}%
			\Sn\pgfmathprintnumber[fixed, precision=\precisionAbsEe, zerofill]{\Mn}%
		}%
		{%
			\ifthenelse{\En=2}%
			{%
				\pgfmathparse{\eval{\Mn*100}}%
				\edef\Mn{\pgfmathresult}%
				\pgfmathparse{int(\precisionAbsEe+1)}%
				\edef\precisionAbsEe{\pgfmathresult}%
				\Sn\pgfmathprintnumber[fixed, precision=\precisionAbsEe, zerofill]{\Mn}%
			}%
			{%
				\ifthenelse{\En=-1}%
				{%
					\pgfmathparse{\eval{\Mn/10}}%
					\edef\Mn{\pgfmathresult}%
					\pgfmathparse{int(\precisionAbsEe+1)}%
					\edef\precisionAbsEe{\pgfmathresult}%
					\Sn\pgfmathprintnumber[fixed, precision=\precisionAbsEe, zerofill]{\Mn}%
				}%
				{%
					\ifthenelse{\En=-2}%
					{%
						\pgfmathparse{\eval{\Mn/10}}%
						\edef\Mn{\pgfmathresult}%
						\pgfmathparse{int(\precisionAbsEe+1)}%
						\edef\precisionAbsEe{\pgfmathresult}%
						\Sn\pgfmathprintnumber[fixed, precision=\precisionAbsEe, zerofill]{\Mn}%
					}%
					{%
						\Sn\pgfmathprintnumber[std, precision=\precisionAbsEe, zerofill]{\Mn} \cdot10^{\En}%
					}%
				}%
			}%
		}%
	}%
}

\tikzstyle{ghostA} = [text=red!70,thick, minimum size=2*(5pt-\pgflinewidth), inner sep=0pt, outer sep=0pt]
\tikzstyle{ghostB} = [text=blue!70,thick, minimum size=2*(5pt-\pgflinewidth), inner sep=0pt, outer sep=0pt]
\tikzstyle{siteA} = [regular polygon, regular polygon sides=3, shape border rotate= 30, draw=red!50,fill=red!20,thick,inner sep=0pt,minimum width=1.5em,font=\footnotesize]
\tikzstyle{siteB} = [regular polygon, regular polygon sides=3, shape border rotate= -30, draw=green!50,fill=green!20,thick,inner sep=0pt,minimum width=1.5em,font=\footnotesize]
\tikzstyle{op} = [regular polygon, regular polygon sides=4, draw=orange!50, fill=orange!20, thick, inner sep=0.2pt, minimum width=1.25em, minimum height=1.5em,font=\footnotesize]
\tikzstyle{opghost} = [regular polygon, regular polygon sides=4, thick, inner sep=0.2pt, minimum width=1.25em, minimum height=1.5em,font=\footnotesize]
\tikzstyle{site} = [circle,draw=blue!50,fill=blue!20,thick,inner sep=0.2pt,minimum width=0.75em,font=\footnotesize]
\tikzstyle{hiddensite} = [circle,draw=white!50,fill=white!20,thick,inner sep=0.2pt,minimum width=1.25em,font=\footnotesize]
\tikzstyle{nosite} = [circle,draw=white,fill=white,thick,inner sep=0.2pt,minimum width=1.25em]
\tikzstyle{ghost} = [font=\footnotesize]
\tikzstyle{intersite} = [regular polygon, regular polygon sides=4, shape border rotate= 45, draw=black!50,fill=black!20,thick,inner sep=0pt,minimum width=1.5em]
\tikzstyle{ld} = [inner sep=1pt, font=\small]
\tikzstyle{unsite} = [circle, outer sep=0pt,inner sep=0.2pt,minimum width=0.75em]